\documentclass[twocolumn,showpacs,floatfix,secnumarabic,amssymb,
amsmath,aps,prl,nobibnotes,superscriptaddress,letter]{revtex4}

\usepackage{revsymb}
\usepackage{graphicx}
\usepackage{dcolumn}
\usepackage{bm}
\usepackage{epstopdf}

\begin{document}

\title{Longitudinal double-spin asymmetry for inclusive jet production in $\vec{p}+\vec{p}$ collisions at $\sqrt{s}$=200 GeV}

\affiliation{Argonne National Laboratory, Argonne, Illinois 60439}
\affiliation{University of Birmingham, Birmingham, United Kingdom}
\affiliation{Brookhaven National Laboratory, Upton, New York 11973}
\affiliation{California Institute of Technology, Pasadena, California 91125}
\affiliation{University of California, Berkeley, California 94720}
\affiliation{University of California, Davis, California 95616}
\affiliation{University of California, Los Angeles, California 90095}
\affiliation{Carnegie Mellon University, Pittsburgh, Pennsylvania 15213}
\affiliation{University of Illinois at Chicago, Chicago, Illinois 60607}
\affiliation{Creighton University, Omaha, Nebraska 68178}
\affiliation{Nuclear Physics Institute AS CR, 250 68 \v{R}e\v{z}/Prague, Czech Republic}
\affiliation{Laboratory for High Energy (JINR), Dubna, Russia}
\affiliation{Particle Physics Laboratory (JINR), Dubna, Russia}
\affiliation{University of Frankfurt, Frankfurt, Germany}
\affiliation{Institute of Physics, Bhubaneswar 751005, India}
\affiliation{Indian Institute of Technology, Mumbai, India}
\affiliation{Indiana University, Bloomington, Indiana 47408}
\affiliation{Institut de Recherches Subatomiques, Strasbourg, France}
\affiliation{University of Jammu, Jammu 180001, India}
\affiliation{Kent State University, Kent, Ohio 44242}
\affiliation{University of Kentucky, Lexington, Kentucky, 40506-0055}
\affiliation{Institute of Modern Physics, Lanzhou, China}
\affiliation{Lawrence Berkeley National Laboratory, Berkeley, California 94720}
\affiliation{Massachusetts Institute of Technology, Cambridge, MA 02139-4307}
\affiliation{Max-Planck-Institut f\"ur Physik, Munich, Germany}
\affiliation{Michigan State University, East Lansing, Michigan 48824}
\affiliation{Moscow Engineering Physics Institute, Moscow Russia}
\affiliation{City College of New York, New York City, New York 10031}
\affiliation{NIKHEF and Utrecht University, Amsterdam, The Netherlands}
\affiliation{Ohio State University, Columbus, Ohio 43210}
\affiliation{Panjab University, Chandigarh 160014, India}
\affiliation{Pennsylvania State University, University Park, Pennsylvania 16802}
\affiliation{Institute of High Energy Physics, Protvino, Russia}
\affiliation{Purdue University, West Lafayette, Indiana 47907}
\affiliation{Pusan National University, Pusan, Republic of Korea}
\affiliation{University of Rajasthan, Jaipur 302004, India}
\affiliation{Rice University, Houston, Texas 77251}
\affiliation{Universidade de Sao Paulo, Sao Paulo, Brazil}
\affiliation{University of Science \& Technology of China, Hefei 230026, China}
\affiliation{Shanghai Institute of Applied Physics, Shanghai 201800, China}
\affiliation{SUBATECH, Nantes, France}
\affiliation{Texas A\&M University, College Station, Texas 77843}
\affiliation{University of Texas, Austin, Texas 78712}
\affiliation{Tsinghua University, Beijing 100084, China}
\affiliation{Valparaiso University, Valparaiso, Indiana 46383}
\affiliation{Variable Energy Cyclotron Centre, Kolkata 700064, India}
\affiliation{Warsaw University of Technology, Warsaw, Poland}
\affiliation{University of Washington, Seattle, Washington 98195}
\affiliation{Wayne State University, Detroit, Michigan 48201}
\affiliation{Institute of Particle Physics, CCNU (HZNU), Wuhan 430079, China}
\affiliation{Yale University, New Haven, Connecticut 06520}
\affiliation{University of Zagreb, Zagreb, HR-10002, Croatia}

\author{B.I.~Abelev}\affiliation{University of Illinois at Chicago, Chicago, Illinois 60607}
\author{M.M.~Aggarwal}\affiliation{Panjab University, Chandigarh 160014, India}
\author{Z.~Ahammed}\affiliation{Variable Energy Cyclotron Centre, Kolkata 700064, India}
\author{B.D.~Anderson}\affiliation{Kent State University, Kent, Ohio 44242}
\author{D.~Arkhipkin}\affiliation{Particle Physics Laboratory (JINR), Dubna, Russia}
\author{G.S.~Averichev}\affiliation{Laboratory for High Energy (JINR), Dubna, Russia}
\author{Y.~Bai}\affiliation{NIKHEF and Utrecht University, Amsterdam, The Netherlands}
\author{J.~Balewski}\affiliation{Indiana University, Bloomington, Indiana 47408}
\author{O.~Barannikova}\affiliation{University of Illinois at Chicago, Chicago, Illinois 60607}
\author{L.S.~Barnby}\affiliation{University of Birmingham, Birmingham, United Kingdom}
% ...

\collaboration{STAR Collaboration}
\noaffiliation

\date{\today}
\begin{abstract}

We report a new STAR measurement of the longitudinal double-spin
asymmetry $A_{LL}$ for inclusive jet production at mid-rapidity in polarized 
$p+p$ collisions at a center-of-mass energy of $\sqrt{s} = 200\,\mathrm{GeV}$.  
The data, which cover jet transverse momenta $5 < p_T <30 \,\mathrm{GeV/c}$, 
are substantially more precise than previous measurements.  They provide significant 
new constraints on the gluon spin contribution to the nucleon spin through the 
comparison to predictions derived from one global fit of polarized deep-inelastic 
scattering measurements.

\end{abstract}

\maketitle

The gluon spin contribution to the nucleon spin, $\Delta G$, has been the focus of experimental and theoretical efforts since polarized deep-inelastic scattering (DIS) experiments found that the quark contribution to the nucleon spin is small~\cite{Ashman:1989ig}.  Insight into $\Delta G$ has been obtained from next-to-leading-order perturbative quantum chromodynamics (NLO pQCD) analyses of the scale dependence of the inclusive spin structure function~\cite{Adeva:1998vw,Anthony:2000fn,Gluck:2000dy,deFlorian:2005mw,Leader:2005ci,Airapetian:2007mh} and from measurements of hadroproduction of pions and photons~\cite{Adams:1991rx}.  Recent semi-inclusive DIS measurements~\cite{Airapetian:1999ib} and hadroproduction measurements of jets and pions have been made~\cite{Abelev:2006uq,Adler:2006bd}, with the latter now being incorporated in NLO pQCD fits~\cite{Hirai:2006sr}.  Despite recent progress, significant uncertainty remains regarding the magnitude and sign of $\Delta{G}$~\cite{Leader:2006xc}.  The inclusive measurements presented here span more than an order of magnitude in partonic momentum fraction ($x$) and are expected to sample a sizable piece
of the total integral $\Delta{G}=\int_0^1{\Delta{g(x)}dx}$.  Comparisons with predictions derived from one global fit~\cite{Jager:2004jh} to deep-inelastic scattering measurements demonstrate the substantial new constraints these results place on $\Delta{G}$.

In this paper, we report a new measurement of the longitudinal double-spin asymmetry $A_{LL}$ for mid-rapidity inclusive jet production in polarized $p+p$ collisions at $\sqrt{s} = 200\,\mathrm{GeV}$ center-of-mass energy,

\begin{equation}
  A_{LL} = \frac{\sigma^{++} - \sigma^{+-}}
{\sigma^{++} + \sigma^{+-}},
\end{equation}

\noindent where $\sigma^{++}$($\sigma^{+-}$) is the differential cross section when the beam protons have equal (opposite) helicities.   We have previously measured the helicity-averaged cross section~\cite{Abelev:2006uq} for transverse momenta ($p_T$) up to $\sim$$50\,\mathrm{GeV/c}$ and it is well described by NLO pQCD evaluations.  Inclusive jet production in the kinematic regime studied here is dominated by gluon-gluon ($gg$) and quark-gluon ($qg$) scattering. Therefore, $A_{LL}$ provides direct sensitivity to gluon polarization~\cite{Jager:2004jh} and the cross-section result motivates the use of NLO pQCD to interpret  our measurements.

The data presented here are extracted from an integrated luminosity of $2\,\mathrm{pb}^{-1}$ recorded in the year 2005 with the STAR detector~\cite{NIM-RHIC} at RHIC. The polarization was measured independently for each of the two counter-rotating proton beams and for each fill using Coulomb-Nuclear Interference (CNI) proton-Carbon polarimeters~\cite{Jinnouchi:2004up}, which were calibrated via a polarized atomic hydrogen gas-jet target~\cite{Okada:2006dd}.  Averaged over RHIC fills,  the luminosity weighted polarizations for the two beams were $52\pm3\%$ and $48\pm3\%$.  The proton helicities were alternated between successive bunches in one beam and between bunch pairs in the other beam. Additionally, the helicity configurations of the colliding beam bunches  were changed between beam fills to minimize systematic uncertainties in the $A_{LL}$ measurement.  Segmented Beam-Beam Counters (BBC)~\cite{BBC:2005}  located up and downstream of the STAR interaction region (IR) measured the helicity dependent relative luminosities, identified minimum bias (MB) collisions,  and served as local polarimeters.

The STAR subsystems used to measure jets are the Time
Projection Chamber (TPC) and the Barrel Electromagnetic Calorimeter
(BEMC)~\cite{NIM-RHIC}.  The TPC provides tracking for charged particles in the 0.5\,T solenoidal magnetic field for pseudo-rapidities $-1.3 \lesssim\eta\lesssim1.3$ and $2 \pi$ in the azimuthal angle $\phi$.  In 2005 the BEMC, covering a fiducial area of  $\phi=2\pi$ and $0<\eta<1$, provided triggering and detection of photons and electrons. 

Events were recorded if they satisfied both the MB condition, defined as a coincidence between east and west BBCs, and either a jet patch (JP) or high tower (HT) trigger. The HT condition required the energy of a single calorimeter tower to be at least $2.6$ (HT1) or $3.6$ (HT2) GeV. The JP trigger fired if the sum of a $\Delta\eta\times\Delta\phi=1\times{1}$ patch of towers, the typical size of a jet, exceeded $4.5$ (JP1) or $6.5$ (JP2) GeV.  Approximately half of the $2.38\times10^6$ jets extracted from the $12\times10^6$ event set originated from the JP2 trigger sample.

Jets were reconstructed using a mid-point cone algorithm~\cite{Blazey:2000qt} with the same parameters as described in Ref.~\cite{Abelev:2006uq}. The algorithm clusters TPC charged track momenta and BEMC tower energies within a cone radius of $R=\sqrt{\Delta\phi^2 +\Delta\eta^2} = 0.4$. Jets were required to have $p_T>5$ GeV/c and point between $\eta=0.2-0.8$ in order to minimize the effects of the BEMC acceptance on the jet energy scale. BBC timing information was used to select events with reconstructed vertex positions within $\sim$60 cm of the center of the detector, ensuring uniform tracking efficiency and matching the conditions used in determining the relative luminosity measurements. Beam background from upstream sources  observed as neutral energy deposits in the BEMC were minimized by requiring the neutral energy fraction of the jet energy (NEF) to be less than 0.8. A minimum NEF of 0.1 was also imposed in order to reduce pile-up effects. Finally, only jets which contained a trigger tower or pointed to a triggered jet patch were considered for analysis.

\begin{figure}[b]
\includegraphics*[height=4cm, width=8.7cm]{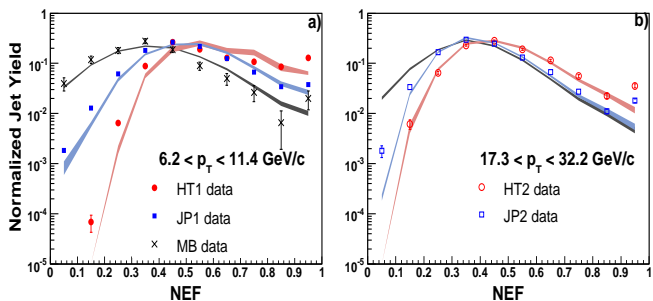}
\caption{\label{fig:NEF} 
Neutral energy fraction of the jet energy for MB (crosses), HT (circles), and JP (squares) data compared
with STAR simulations for two jet $p_T$ bins, (a) $6.2 < p_T < 11.4$ GeV/c and (b) $17.3 < p_T < 32.2$ GeV/c. The statistical uncertainties are represented as error bars for the data points and bands for the simulations. (Color available on-line)}
\end{figure}

Figure~\ref{fig:NEF}  compares the NEF spectra for MB, HT, and JP triggered jets
from  data and simulations. Monte Carlo events were generated using  PYTHIA 6.205~\cite{Sjostrand:2001yu} with parameters adjusted to CDF 'Tune A' settings~\cite{Field:2005sa} and processed through  the STAR detector response package based on GEANT 3 ~\cite{geant}.  The shapes of the data distributions are sufficiently reproduced by the simulations for the purpose of estimating systematic errors. In contrast to the calorimeter triggers, the mean and shape of the MB distribution is relatively stable as a function of jet $p_T$. The HT jets, and to a lesser extent the JP jets, show a strong bias towards higher NEF at low $p_T$ which diminishes for higher jet $p_T$. The enhancement of jets near NEF$\approx$1 in the data compared to simulation is consistent with contributions from beam background, as discussed above. 

\begin{figure}[t]
\includegraphics*[height=4cm,width=8.5cm]{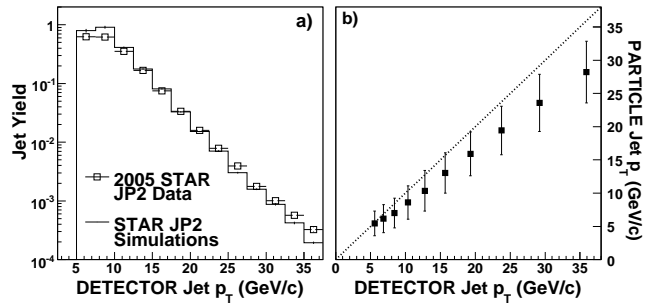}
\caption{\label{fig:Reso2D} 
(a) The raw detected jet yield in data (points) compared with the STAR Monte Carlo simulations.
(b) Correlation between the reconstructed jet transverse momenta at the {\scshape particle} and {\scshape detector} levels. The points indicate the means and the vertical error bars show the r.m.s. widths of the associated {\scshape particle} jet distributions within the {\scshape detector} jet bins. The dashed line represents the condition when the {\scshape particle} and {\scshape detector} jet $p_T$ values are equal.}
\end{figure}

We present the inclusive jet $A_{LL}$ measurement, not as a function of the measured transverse momentum ({\scshape detector} jet $p_T$), but instead corrected to reflect the jet energy scale before interaction with the STAR detector ({\scshape particle} jet $p_T$).  This correction was carried out by applying the same jet reconstruction algorithm to the simulated event samples at both the {\scshape particle} and {\scshape detector} levels. {\scshape Particle} jets are composed of stable, final-state particles which result from the fragmentation and hadronization of the scattered partons and remnant protons. {\scshape Detector} jets consist of the reconstructed TPC tracks and BEMC tower energies in simulated events that pass the same trigger conditions that were placed on the data. As shown in  Fig.~\ref{fig:Reso2D}(a) the jet yield is a rapidly falling function of jet $p_T$.  This effect combined with the jet $p_T$ resolution of about 25$\%$ results in  a shift of the inclusive {\scshape detector} jet $p_T$ distribution to larger values as shown in Fig.~\ref{fig:Reso2D}(b).  For $p_{T} > 10 \, \mathrm{GeV/c}$ the reconstructed {\scshape detector} jet $p_T$ is on average a factor 1.22 larger than the {\scshape particle} jet $p_T$.  This shift varies only slightly with trigger and underlying partonic process ($gg$ vs. $qg$ vs. $qq$).  The dominant uncertainty in the jet $p_T$ values arises from the $\pm$\,4\% uncertainty in the jet energy scale, but we also account for  the subsample dependence in the {\scshape detector} to {\scshape particle} jet conversion.  The agreement between data and simulations is best at high $p_T$ (as shown in Fig.\ref{fig:NEF})  where the correction, proportional to $p_T$, is largest. The uncertainty in the {\scshape particle} jet $p_T$ slope results in a $\pm2.5\%$  error on the $p_T$ scale. 

The asymmetry $A_{LL}$ was evaluated according to,
\begin{equation}
A_{LL}=
  \frac{
       \sum \left(P_1 P_2\right) \left( N^{++} - R N^{+-} \right)
  }{
       \sum \left(P_1 P_2\right)^2 \left( N^{++}  + R N^{+-} \right)
  },
\end{equation}
in which $P_{1,2}$ are the measured beam polarizations, $N^{++}$ and $N^{+-}$ denote the inclusive jet yields for equal and opposite proton beam helicity configurations, and $R$ is the measured relative luminosity.  Each sum is over 10  to 30 minute long runs, a period much shorter than typical time variations in critical quantities such as $P_{1,2}$ and $R$.  Typical values of $R$ range from 0.85 to 1.2 depending on fill and bunch pattern.  

\begin{figure}[htp!]
\includegraphics*[scale=0.45]{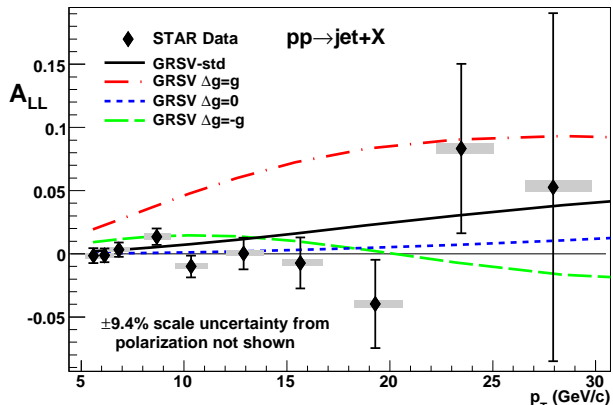}
\vspace{-0.8cm}
\caption{\label{fig:aLL}  Longitudinal double-spin asymmetry $A_{LL}$ for inclusive jet production at $\sqrt{s}$ = 200 GeV versus jet $p_T$.  The points show results for {\scshape particle} jets with statistical error bars, while the curves show predictions for NLO parton jets from one global analysis~\cite{Jager:2004jh}.  The gray boxes indicate the systematic uncertainties on the measured $A_{LL}$ values (vertical) and in the corrections to the measured jet $p_T$ and the conversion between {\scshape particle} jet and NLO parton jet $p_T$ (horizontal). (Color available on-line.)}
\end{figure}

Figure~\ref{fig:aLL} shows the results for inclusive jet $A_{LL}$ versus jet $p_T$ corrected for detector response to the particle level. The vertical error bars show the statistical uncertainties. The height of the gray band on each data point  indicates the total systematic uncertainty on $A_{LL}$ while the width reflects the systematic uncertainty on jet $p_T$. An overall 9.4\% scale error due to the uncertainty in the RHIC CNI polarimeter calibration is not included in the systematic error shown in Fig.~\ref{fig:aLL}.  The present results are in good agreement ($\chi^2/ndf = 7.3/6$) with our previous measurements of $A_{LL}$~\cite{Abelev:2006uq}.  The combined statistical and systematic asymmetry uncertainties are reduced by a factor of 4 and the $p_T$ coverage is nearly doubled, extending up to 30 $\mathrm{GeV/c}$. 

The curves shown in Fig.~\ref{fig:aLL} are  NLO pQCD evaluations for $A_{LL}$ based on different polarized parton distribution functions~\cite{Jager:2004jh}. The curve labeled GRSV-std uses the best fit to the inclusive DIS data ($\Delta{g}(x,Q^2_0)=0.24$) at the initial scale $Q_{0}^{2}=0.4\,\mathrm{GeV^2/c^2}$ of the parton parametrizations in Ref.~\cite{Gluck:2000dy}.  The other curves correspond to maximally  positive ($\Delta{g}(x,Q^2_0)=g$), maximally negative  ($\Delta{g}(x,Q^2_0)=-g$) and vanishing ($\Delta{g}(x,Q^2_0)=0$) gluon polarization.  The calculations were performed with factorization and renormalization scales $\mu_F=\mu_R=p_T$.  The $A_{LL}$ values for the GRSV-std and $\Delta{g}(x,Q^2_0)=g$ cases change by less than 20\% for variations in the scale from $p_T/2$ to $2p_T$.  The calculations are performed for jets composed of NLO partons which do not include effects due to hadronization and the underlying event.  This difference, estimated from simulation studies,  causes a  $^{+4}_{-0}$\% systematic shift in jet $p_T$ between {\scshape particle} and NLO parton jets. 

The leading systematic error contribution to the  $A_{LL}$ measurement arises from trigger and reconstruction effects which cause the asymmetry to differ for {\scshape particle} and {\scshape detector} jets.  The shift in jet energy scale results in the smearing of the {\scshape particle} jet $A_{LL}$  across the detected jet $p_T$ bin, an effect which is largely accounted for by the correction of measured {\scshape detector} jet to {\scshape particle} jet $p_T$ values.  The calorimeter triggers, designed to select a subset of all minimum bias events, change the natural distribution of $qq$, $qg$ and $gg$ events which comprise the
inclusive measurement. The consequence  of this change for  $A_{LL}$ also depends on the true value of the gluon helicity distribution. Therefore the systematic error due to both triggering  and reconstruction bias was estimated from the jet asymmetries calculated within the simulation framework for GRSV-std, $\Delta{g}(x,Q^2_0)=0$, and $\Delta{g}(x,Q^2_0)=-g$ scenarios.  The $\Delta{g}(x,Q^2_0)=g$ scenario, shown in Fig.~\ref{fig:aLL}, is not consistent with our data and therefore was not included in the estimates. The maximum positive and negative differences for the distributions were selected at each $p_T$ bin.  Other systematic uncertainties include effects from relative luminosities $(9\times10^{-4})$,  beam background $(7\times10^{-4})$ and non-longitudinal beam polarization components at the STAR IR $(3\times10^{-4})$.  Parity violating single-spin asymmetries in the data were found to be consistent with zero, $< 0.2$ standard deviation, as expected, given that parity violating physical processes are predicted to be negligible at the current level of statistics.  

\begin{figure}[htp!]
\includegraphics*[scale=0.45]{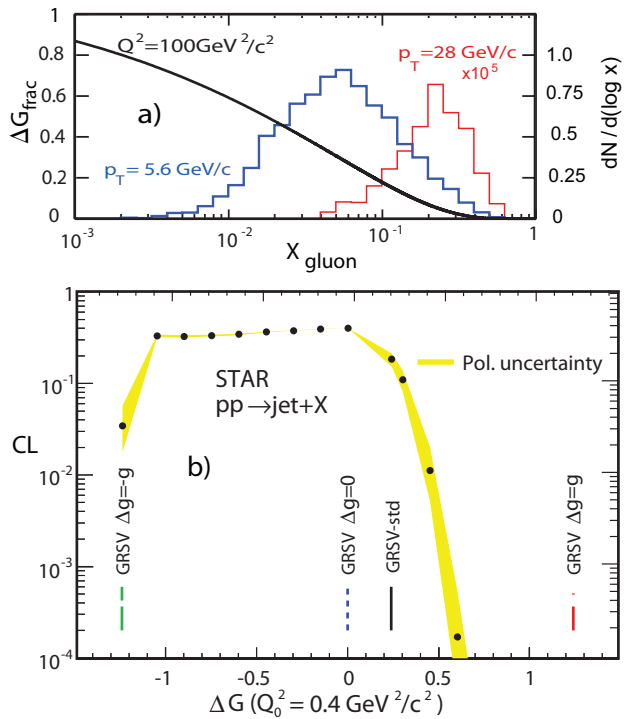}
\caption{\label{fig:cl} 
(a) The solid curve represents the fraction of $\Delta{G}$ for GRSV-std that has $x>x_{min}$ for scale $Q^2=100\,\mathrm{GeV^2/c^2}$. The histograms show the $x_{gluon}$ sampled in the lowest and highest jet $p_T$ bins. (b) Confidence levels for several gluon polarization distributions, characterized by their $\Delta{G}$ at an input scale of 0.4$\,\mathrm{GeV^2/c^2}$~\cite{Gluck:2000dy,Jager:2004jh,SVpriv} (Color available on-line).}
\end{figure}

Figure~\ref{fig:cl}a illustrates the gluon $x$ range accessed in a low and high $p_T$ data bin.  Note that the sampled $\Delta g(x)$ distributions depend on the magnitude and momentum-dependence of the gluon polarization and may have very different shapes. The smooth curve represents the corresponding fraction of $\Delta G$  sampled for $x_{min}=x_{gluon}$ in the GRSV-std scenario at a scale of $Q^2=100$ GeV$^2/c^2$ which is typical for the present data.  The measurements presented here provide sensitivity to $\sim$50$\%$ of $\Delta{G}$. 
 
To quantify the impact of the new data on $\Delta G$, the measured $A_{LL}$ values have been compared to predictions within the GRSV framework~\cite{Gluck:2000dy,Jager:2004jh,SVpriv} in which the polarized parton distributions were re-fit assuming $\Delta{G}$ is constrained to a series of values spanning the full range $-g\leq\Delta{g}(x,Q^2_0)\leq{g}$.  Figure~\ref{fig:cl}b shows the confidence levels (C.L.) found from comparisons with this inclusive jet data. The correlations among the systematic uncertainties for various jet $p_T$ have been included in the C.L. calculations. 

We find that the presented data exclude fits of $\Delta{G} > 0.33$ at a scale of 0.4 GeV$^2$/c$^2$ with at least 90\% C.L. and the $\Delta{g}(x,Q^2)=-g$ scenario is excluded at the 94\% level.  As discussed in Ref.~\cite{Gluck:2000dy},  the GRSV-std fit to the existing DIS world data corresponded to a $\Delta{G}(Q^2_0=0.4)= 0.24$ with a range of $-0.45<\Delta{G}<0.7$ allowed with a $\chi^2$ variation of 1. Although these conclusions are dependent on the functional form for the gluon polarization defined in the GRSV framework, the constraints placed by our data on the slice of $\Delta{G}$ between x=0.02-0.2 are significant and will exclude additional PDF's which have a $\Delta{G}=\int_{0.02}^{0.2}{\Delta{g(x)}dx}$ contribution larger than GRSV-std in this x region. 

In summary, we report new measurements of the longitudinal
double-spin asymmetry $A_{LL}$ for inclusive jet production at
mid-rapidity in polarized $p+p$ collisions at $\sqrt{s}$ = 200 GeV
with coverage in jet transverse momentum up to $30 \,\mathrm{GeV/c}$
and improved precision compared to previous measurements. The data
have been compared to predictions of $A_{LL}$ within the GRSV framework,  
a commonly used set of polarized parton distribution functions, constraining 
$\Delta{G}(Q^2_0)$ to less than $65\%$ of the proton spin with $90\%$ confidence.
A global analysis of DIS and RHIC data is needed to realize the full impact of 
these results on the shape and magnitude of $\Delta g(x,Q^2)$.

The authors thank W.~Vogelsang and M.~Stratmann
for providing calculations and discussion.
We thank the RHIC Operations Group and RCF at BNL, and the
NERSC Center at LBNL for their support. 
This work was supported
in part by the Offices of NP and HEP within the U.S. DOE Office 
of Science; the U.S. NSF; the BMBF of Germany; CNRS/IN2P3, RA, RPL, and
EMN of France; EPSRC of the United Kingdom; FAPESP of Brazil;
the Russian Ministry of Education and Science; the Ministry of
Education and the NNSFC of China; IRP and GA of the Czech Republic,
FOM of the Netherlands, DAE, DST, and CSIR of the Government
of India; Swiss NSF; the Polish State Committee for Scientific 
Research; SRDA of Slovakia, and the Korea Sci. \& Eng. Foundation.

\bibliography{basename of .bib file}

\break
\begin{table}
\caption{\label{table:asy} $A_{LL}$ for inclusive jet production in jet $p_T$ bins
with statistical (stat) and systematic (sys) uncertainties. Both the average measured  
jet $p_T$, $\langle p_T^M \rangle$,  and the final shifted jet $p_T$  with associated systematics are reported. }
\begin{ruledtabular}
\begin{tabular}{llll}
 $\langle p_T^M \rangle $  & $\langle p_T \rangle$+sys.-sys. &  $A_{LL}$ $\pm$ stat +sys-sys   \\
 (GeV/c) &  (GeV/c) &  (x10$^{-3}$)  \\
\hline
 5.58 & 5.60 +0.36 -0.41  & -1.51 $\pm$ 5.92 +2.11 -2.11  \\
 6.84 & 6.14 +0.46 -0.48  & -1.16 $\pm$ 5.44 +1.77 -1.90 \\
 8.38 & 6.83 +0.53 -0.54  &  3.27 $\pm$ 5.59 +1.61 -1.86  \\
10.26 & 8.67 +0.65 -0.61  & 13.51 $\pm$ 6.53 +2.91 -2.15 \\
12.57 & 10.34 +0.83 -0.76  & -10.01 $\pm$ 8.58 +2.60 -1.71 \\
15.41 & 12.89 +0.96 -0.83  & 0.19 $\pm$  12.59 +2.77 -1.78 \\
18.90 & 15.65 +1.09 -0.88  & -7.30 $\pm$ 20.21 +3.09 -2.22\\
23.20 & 19.30 +1.31 -1.02  & -39.69 $\pm$ 34.84 +2.99 -2.99 \\
28.39 & 23.48 +1.59 -1.19  & 83.17 $\pm$  67.03 +4.10 -4.10 \\
34.81 & 27.94 +1.88 -1.37  & 52.70 $\pm$  137.65 +5.74 -3.73\\
\end{tabular}
\end{ruledtabular}
\end{table}

\end{document}